\documentclass[journal,twoside,web]{ieeecolor}
\usepackage{tuffc}
\usepackage{cite}
\usepackage{amsmath,amssymb,amsfonts}
\usepackage{algorithmic}
\usepackage{graphicx}
\usepackage{upgreek}
\usepackage{textcomp}
\usepackage{wrapfig,colortbl}
\definecolor{abstractbg}{rgb}{1,0.969,0.914}
\def\BibTeX{{\rm B\kern-.05em{\sc i\kern-.025em b}\kern-.08em
    T\kern-.1667em\lower.7ex\hbox{E}\kern-.125emX}}
\begin{document}
\title{Terahertz molecular frequency references: theoretical analysis of optimal instability}
\author{W. F. McGrew, \IEEEmembership{Member, IEEE}, James Greenberg, \IEEEmembership{Member, IEEE}, Keisuke Nose, Brendan M. Heffernan, \IEEEmembership{Member, IEEE}, and Antoine Rolland, \IEEEmembership{Member, IEEE}
\thanks{Manuscript received XXX; accepted XXX. This work was funded by Imra America, Inc. }
\thanks{The authors are with Imra America, Inc., Boulder Research Labs, Longmont, CO (email: wmcgrew@imra.com). }}

\IEEEtitleabstractindextext{%
\fcolorbox{abstractbg}{abstractbg}{%
\begin{minipage}{\textwidth}\rightskip2em\leftskip\rightskip\bigskip
\begin{wrapfigure}[12]{r}{2.0in}%
\hspace{-3pc}\includegraphics[width=2.0in]{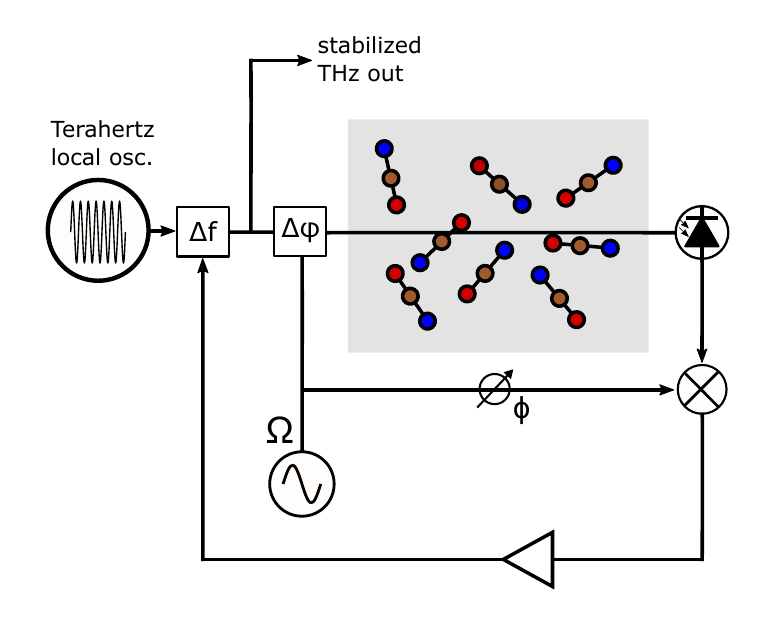}
\end{wrapfigure}%

\begin{abstract}
We report a comprehensive theoretical analysis of the instability achievable by using phase modulation spectroscopy to lock a terahertz local oscillator to an absorptive reference consisting of the rotational transition of molecules at room temperature. We find that the signal-to-noise ratio of the THz detector provides the limitation to the instability that can be achieved and analyze a number of viable candidate molecules, identifying several as being of particular interest, including OCS and HI. We find that a one-second instability in the $10^{-13}$ decade is achievable for molecules confined to waveguide, while instability at the $10^{-14}$ level can be reached for molecules in free space. We also present calculations of the intermodulation effect for spectroscopy taking place far outside the quasistatic regime and find that this source of noise presents constraints on which THz local oscillators are appropriate to be used for such frequency references.
\end{abstract}

\bigskip
\end{minipage}}}

\maketitle

\section{Introduction}
\label{sec:introduction}
\IEEEPARstart{T}{he} demand for next-generation, portable commercial frequency references has motivated a surge of research in a number of metrological platforms, focusing especially on optical references such as iodine \cite{Roslund2024}, acetylene \cite{Droste:24}, and the rubidium two-photon transition \cite{Beard2024}, among others. However, the increasing maturation of technologies operating at terahertz frequencies \cite{Tonouchi2007} motivates the development of standards in this range (0.1 to 10 THz). Compared with their optical counterparts, THz references have the promise of division down to microwave frequencies without the necessity of a self-referenced frequency comb, while still potentially outperforming otherwise comparable microwave standards due to the much higher transition frequency of THz standards. Furthermore, the ongoing development of next-generation, ``6G," wireless communication protocols \cite{Chowdhury2020} stands to benefit from the ready availability of native THz references.

With such motivations, a number of laboratories have recently investigated molecular rotational THz transitions as frequency references, exploiting N$_2$O \cite{Greenberg2024b}, OCS \cite{Greenberg2024, Wang2018}, and CO \cite{Rothbart2024}, and such experiments have achieved Allan deviations as low as $1.7 \times 10^{-12}/\sqrt{\tau}$ (N.B.: the modified Allan deviation, quoted in \cite{Greenberg2024}, is smaller by factor of $\sqrt{2}$). To our knowledge, a comprehensive theory of such frequency references is not available in the literature, and this work aims to provide such a treatment for several molecules of interest, in both waveguide and free space configurations.

\begin{figure}[!t]
\centerline{\includegraphics[width=\columnwidth]{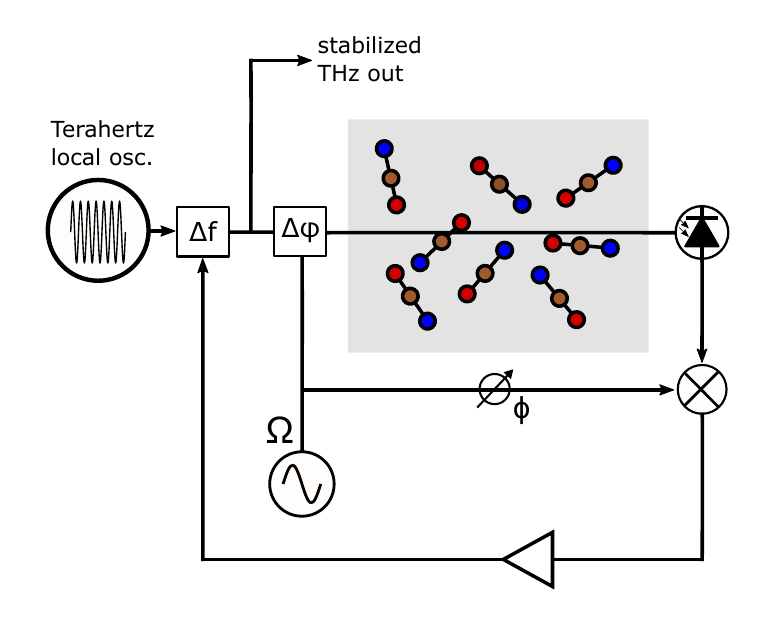}}
\caption{The basic experimental setup analyzed in this work. A terahertz local oscillator (for instance a down-converted optical source, upconverted microwave source, or native THz source) is stabilized to a molecular transition by phase modulation spectroscopy. A terahertz detector measures the transmitted electromagnetic radiation, and this signal is demodulated, amplified, and integrated to provide feedback to the local oscillator. The stabilized signal is output.}
\label{fig1}
\end{figure}

This article will focus on continuous probing schemes based on phase-modulation spectroscopy \cite{Supplee1994}. There are three basic components in such a scheme, as illustrated in Fig. \ref{fig1}. First, a THz local oscillator is chosen; this can be an optical reference (such as a dual-wavelength Brillouin laser \cite{Heffernan2024} or a dissipative Kerr soliton comb \cite{Zhang2019, Tetsumoto2021}) downconverted by an ultrahigh bandwidth photomixer like a unitraveling carrier photodiode (UTC-PD) \cite{Ishibashi2020}, an up-converted microwave signal \cite{Maiwald2003, Yi2021}, or an intrinsically THz source (such as a resonant tunneling diode \cite{Asada2021} or a quantum cascade laser \cite{Vitiello2015}). The second component is propagation of the synthesized THz waves through the molecular absorptive medium, which could be carried out in either waveguide or freespace. As will be shown in this work, the molecular pressure, beam configuration, field intensity, and so forth are important parameters for the achievable instability. Finally, after interaction with the molecules, a detector is used to transduce frequency fluctuations into a voltage, which can be integrated and fed back to a convenient actuator for the local oscillator. 

We present a generalized method for predicting the experimentally observed fractional frequency instability. To validate our model, we focus on our recent experimental implementation of a dual-wavelength Brillouin laser photomixed by UTC-PD and detected by a zero-bias Schottky barrier diode (SBD). Then, we explore operating conditions and molecule choice to optimize for the lowest instability. To narrow the scope of this optimization, we only considered the SBD from our experiment and emphasize that the conclusions of this paper may differ with other terahertz detectors. Furthermore, we focus our analysis to the range of frequencies at the water window near 300 GHz, with practical communications applications in mind \cite{Schneider2012}.


\begin{table*}[!t]
\arrayrulecolor{subsectioncolor}
\setlength{\arrayrulewidth}{1pt}
{\sffamily\bfseries\begin{tabular}{lp{6.75in}}\hline
\rowcolor{abstractbg}\multicolumn{2}{l}{\color{subsectioncolor}{\itshape
Highlights}{\Huge\strut}}\\
\rowcolor{abstractbg}$\bullet$ & \textbf{Novelty:} This work introduces a comprehensive theoretical framework to evaluate the instability of terahertz (THz) molecular frequency references using phase modulation spectroscopy, identifying optimal molecules and configurations for stability improvement.\\
\rowcolor{abstractbg}$\bullet${\large\strut} & \textbf{Main Results:} The study demonstrates that OCS molecules confined to waveguides achieve instabilities below $10^{-12}$ at one second, while free-space setups can achieve instabilities an order of magnitude lower.\\
\rowcolor{abstractbg}$\bullet${\large\strut} & \textbf{Importance:} These findings pave the way for high-performance, compact THz frequency references, essential for advancing metrological applications, 6G wireless communications, and next-generation molecular clocks.\\[2em]\hline
\end{tabular}}
\setlength{\arrayrulewidth}{0.4pt}
\arrayrulecolor{black}
\end{table*}

\section{Model and validation}

To model the expected instability achieved by THz frequency references, it is first necessary to calculate the error signal that can be achieved with the given experimental architecture. The THz beam is synthesized through photomixing of two laser beams, one of which is phase modulated (by, for instance, passing the beam through an electro-optic modulator). The electric field emitted is given by,

\begin{equation}
    E(t) = E_0 e^{i(2\pi\nu t + \beta \sin 2\pi \Omega t)}, \label{modfield}
\end{equation}

\noindent where $E_0$ is the electric field amplitude, $\nu$ is the (THz-scale) frequency of the carrier, $\beta$ is the phase modulation index, and $\Omega$ is the frequency of the modulation. Exploiting the Jacobi-Anger expansion using Bessel functions, we can expand this as,


\begin{equation}
    \begin{split}
        E(t) &= E_0 e^{i 2\pi\nu t} \left[ J_0(\beta) + 2iJ_1(\beta)\sin{(2\pi\Omega) t} \right. \\
        &\left. \qquad\qquad\qquad + 2 J_2(\beta) \cos{2 (2\pi \Omega) t} + ... \right] \\
        &= E_0 \sum_{n=-\infty}^\infty \left( J_n(\beta)e^{2\pi i(\nu+n\Omega)t} \right) .
    \end{split}
\end{equation}

When the light is passed through an absorptive medium, the transmission signal can be characterized by a complex transmission function, 

\begin{equation}
    T(\nu) = \exp\left(-\delta(\nu)-i\phi(\nu)\right),
    \label{transeq}
\end{equation}

\noindent where $\delta(\nu)$ is the decrease of electric field intensity as a function of frequency, and $\phi(\nu)$ is the phase shift. For this work, we assume a Lorentzian profile,

\begin{equation}
    \delta(\nu) = \frac{A\Gamma^2}{(\nu-\nu_0)^2 + \Gamma^2}, \label{abseq}
\end{equation}

\noindent where $\Gamma$ is the half-width at half-maximum in Hz, $A$ is the maximum absorption (expressed as the ratio of electric field after to that before), and $\nu_0$ is the line center. The phase function, determined by the Kramers-Kronig relations, is then,

\begin{equation}
    \phi(\nu) = \frac{A\Gamma(\nu-\nu_0)}{(\nu-\nu_0)^2 + \Gamma^2}. \label{phseq}
\end{equation}

\noindent After propagation through the absorptive medium, the transmitted signal is then given by,


\begin{equation}
\begin{split}
     E_T=E_0 \sum_{n=-\infty}^\infty \left( J_n(\beta) T(\nu+n\Omega)e^{2\pi i(\nu+n\Omega)t} \right) .
      \label{eleceq}
\end{split}
\end{equation}

For measurements taking place in waveguide, the transmitted electric field amplitude is multiplied by $10^{(- (\alpha/10) L/2)}$, where $L$ is the waveguide length, and $\alpha$ is the waveguide power loss coefficient, which can be of the order of 0.1 dB/cm or more.

The detector measures the intensity, $P_T \propto |E_T|^2$ and emits an electrical signal, as determined by the responsivity of the detector. This signal is then mixed with a signal of the form $\sin(\Omega t - \psi)$, where $\psi$ is a tunable demodulation phase, and a low-pass-filter removes all components that are not small with respect to $|\omega-\Omega|$. For the simulation, a software low-pass filter is used to imitate these electrical components, and the demodulation phase $\psi$ is scanned to numerically optimize the slope of the error signal at $\nu_0$. This procedure numerically reproduces lock-in detection of the phase modulation. In order to calculate this error signal, it is necessary to model the molecular lineshape parameters, $A$ and $\Gamma$. 

\subsection{Molecular line shape}

Molecular broadening mechanisms for rotational transitions of molecules are treated comprehensively in Chapter 13 of \cite{Townes1975}. Below, we concisely summarize the relevant physics. Note that this section employs Gaussian-cgs units.

The first broadening mechanism to be treated is pressure broadening, arising from decohering collisions between gas molecules. The pressure broadening amounts to $\Delta\nu_P = b P$, for gas pressure, $P$. Self-broadening coefficients are generally greater than coefficients for intermolecular collisions, and in practice it is straightforward to reduce the background pressure well below the partial pressure for the molecule of choice, so we expect collisions with background gases to be negligible. Self-broadening coefficients, $b$, have generally been measured with reasonably small uncertainties for most molecules of interest for THz frequency references and range from a few MHz/Torr, to nearly 100 MHz/Torr, for large molecules with significant dipole moments. 

For measurements taking place in waveguide, collisions with the walls lead to decoherence. Assuming the waveguides are long with respect to the cross-sectional width, and further assuming each wall collision is decohering, we can derive from the kinetic theory of gases,

\begin{equation}
    \Delta\nu_\text{wall} = \frac{a}{V} \left(\frac{k_B T}{8 \pi^3 m} \right) ^\frac{1}{2},
\end{equation}

\noindent where $a$ is the internal surface area of the waveguide, $V$ is the volume enclosed, $k_B$ is Boltzmann's constant, $T$ is the temperature of the gas, and $m$ is the mass of a molecule. This broadening mechanism is independent of the waveguide length.

For measurements taking place in free space, the line is further broadened by transit-time effects, due to the two-dimensional thermal velocity transverse to the molecular beam \cite{Demtroder2014}. For a Gaussian THz beam, this broadening has a Gaussian lineshape with a half-width at half-maximum (HWHM) of

\begin{equation}
    \Delta\nu_\text{trans} = \frac{1}{w_0} \left(\frac{ k_B T \ln 2}{2m}\right),
\end{equation}

\noindent where $w_0$ is the beam waist. Since we are interested in the error signal near to resonance for the spectroscopic signal, it is reasonable to treat Gaussian broadening mechanisms as Lorentzian with HWHM corresponding to the Gaussian HWHM.

All of the above broadening mechanisms are caused by an approach to thermal equilibrium in $J$ states sampled by the beam (either through collisions mixing quantum states or by molecules leaving the beam and being replaced with thermal molecules). Therefore, these mechanisms add linearly, yielding $\Delta\nu_\text{therm} = \Delta\nu_P + \Delta\nu_\text{wall (trans)} = 1/2\pi \tau_\text{therm}$, for molecules in waveguide (free space). $\tau_\text{therm}$ is the time scale for thermal decoherence of the $J$ state.

The next broadening mechanism to be considered, saturation by the probe beam, requires care of treatment, as it is the only mechanism addressed here that decreases the integrated line strength. It is characterized by the saturation parameter, 

\begin{equation}
    S = \frac{32 \pi^3 |\mu_{01}|^2 I \tau_\text{therm}^2}{3ch^2},
\end{equation}

\noindent where $|\mu_{01}|^2 = \mu^2 \frac{J+1}{2J+1}$ is the transition dipole moment for the rotational transition $J+1 \leftarrow J$ for a molecule of dipole moment $\mu$, $I$ is the light intensity, $c$ is the speed of light in vacuum, and $h$ is Planck's constant. Saturation causes the Lorentzian line width to be increased by a factor of $\sqrt{1 + S}$ and the absorption constant on resonance to be decreased by a factor of $(1+S)^{-1}$. Including only these broadening mechanisms, the loss coefficient can be expressed as,

\begin{equation}
\begin{split}
    2\delta(\nu) &= \frac{8\pi^2 N F |\mu_{01}|^2 \nu^2}{3 c k_B T L} \frac{\Delta\nu_\text{therm}}{(\nu-\nu_0)^2 + \Delta\nu_\text{therm}^2 (1 + S)} \\
    &= \frac{8\pi^2 N F |\mu_{01}|^2 \nu^2}{3 c k_B T L \sqrt{1+S}} \frac{\Delta\nu_\text{therm}\sqrt{1+S}}{(\nu-\nu_0)^2 + \Delta\nu_\text{therm}^2 (1 + S)},
    \label{satcoll}
\end{split}
\end{equation}

\noindent where $N$ is the number of molecules per cm$^3$ (density), $F$ is the fraction of those molecules in the lower state of interest, and $L$ is the path length. The second line of \eqref{satcoll} makes clear that the integrated line strength is invariant for a changing $\Delta\nu_\text{therm}$ (and fixed $S$) but decreases for increasing $S$.

The final broadening mechanism arises due to the Doppler effect. In a thermal gas, this leads to a Gaussian lineshape with a HWHM of

\begin{equation}
    \Delta\nu_\text{Dopp} = \frac{\nu}{c} \sqrt{\frac{2 k_B T \ln 2}{m}}.
\end{equation}

This is added in quadrature with the linewidth of \eqref{satcoll}, yielding a total linewidth, $\Delta\nu_\text{tot}^2 = \Delta\nu_\text{therm}^2 (1 + S) + \Delta\nu_\text{Dopp}^2$, and an absorption coefficient of,

\begin{equation}
    2\delta(\nu) = \frac{8\pi^2 N F |\mu_{01}|^2 \nu^2}{3 c k_B T L \sqrt{1+S}} \frac{\Delta\nu_\text{tot}}{(\nu-\nu_0)^2 + \Delta\nu_\text{tot}^2}.
\end{equation}

Using the parameters defined in \eqref{abseq}, we have

\begin{equation}
\begin{split}
    \Gamma &= \Delta\nu_\text{tot}, \\
    A &= \frac{4\pi^2 N F |\mu_{01}|^2 \nu^2}{3 c k_B T L \Delta\nu_\text{tot} \sqrt{1+S}}.
\end{split}
\end{equation}

\noindent From these parameters, the phase function, $\phi(\nu)$, can immediately be derived by using \eqref{phseq}.

\subsection{Instability arising from THz detector SNR}

The signal-to-noise ratio (SNR) from the THz detector is a key source of instability for THz frequency references. This source of instability is given by,

\begin{equation} \label{detnoise}
    \sigma = V_\text{noise} / \left(\sqrt{2} \nu_0 \frac{\partial e(\nu)}{\partial\nu} \right) ,  
\end{equation}

\noindent where $V_\text{noise}$ is the root-mean-squared voltage fluctuation from the detector, and $\frac{\partial e(\nu)}{\partial\nu}$ is the slope of the error signal. Research at IMRA has made use of THz zero-bias SBDs. A key noise source is the detector's flicker noise (1/f noise), which increases with input power \cite{Hesler2007}. However, due to the very high (40 GHz) bandwidth of the SBD, it is straightforward to increase the modulation frequency to the point that this noise source can be made as small as desired, and calculation indicates that increasing modulation frequency has no pernicious influence on the slope of the spectroscopic error signal. In practice, a modulation frequency of $>10$ MHz has been found to be adequate to suppress this source of instability.

For high modulation frequency, the dominant noise sources are the shot and thermal noise of the SBD \cite{Trippe1986}, which contributes a current noise power spectral density of

\begin{equation} \label{thermshoteq}
    S_C = \frac{2eCR_j^2}{(R_s + R_j)^2} + \frac{4 k_B T R_s}{(R_s + R_j)^2},
\end{equation}

\noindent where $e$ is the elementary charge, $C$ is the SBD current, $R_j \approx 1$ k$\Upomega$ is the junction resistance, and $R_s \approx 20$ $\Upomega$ is the series resistance. The SNR from the Schottky can be increased by operating at a higher THz power, but saturation of the SBD leads eventually to the relative decrease of signal relative to the noise. Departure from square law operation has been measured to occur at a few 10 $\upmu$W, and optimum SNR is attained at a few 100 $\upmu$W. By measuring the responsivity and $1/f$ noise as a function of input power (see the Appendix), an empirical detector model was constructed which allows us to model the detector response in volts, and $V_\text{noise}$. 

\begin{figure}[!t]
\centerline{\includegraphics[width=\columnwidth]{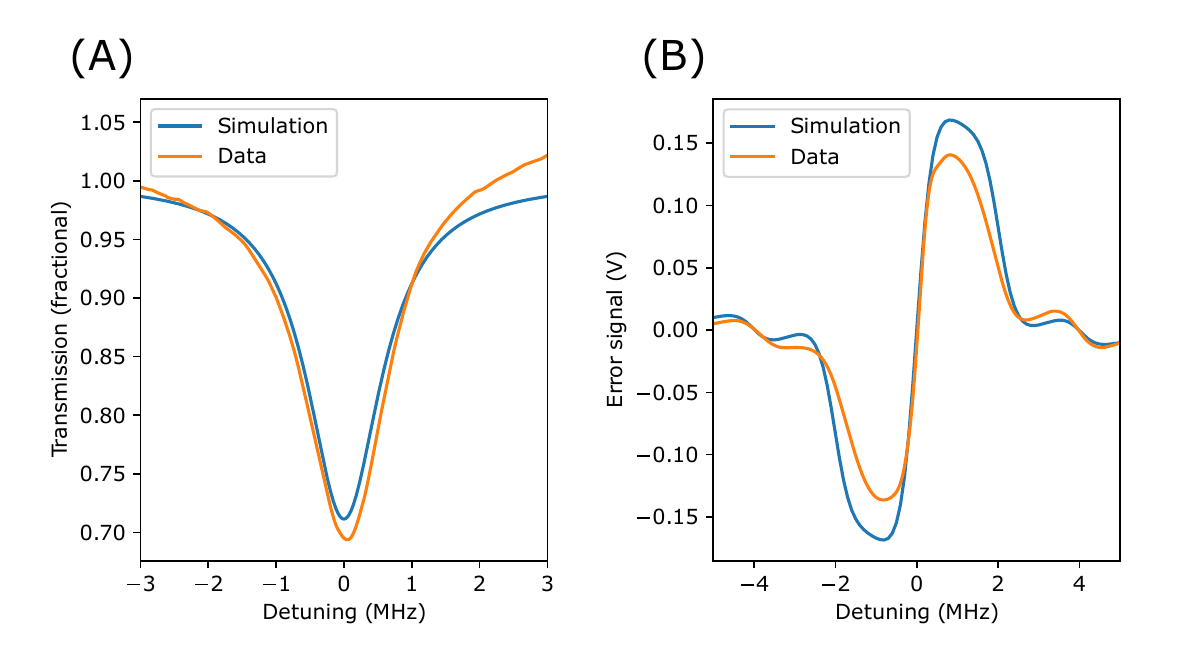}}
\caption{Comparison of theoretical simulation and measured \cite{Greenberg2024} signal for (A) transmission of unmodulated THz radiation, and (B) the error signal measured after demodulation. Asymmetries in the measured absorption and error signals are attributed to variation in the intensity of the THz radiation as the frequency was swept, by changing the frequency driving an acousto-optic modulator.}
\label{fig2}
\end{figure}

Using \eqref{eleceq} and the detector model, the error signal can be calculated ab initio and compared to experimental measurement. Fig \ref{fig2} shows excellent agreement between our calculation of the absorption and error signals, with the experimental values reported in \cite{Greenberg2024} for the $J'=26 \leftarrow J''=25$ transition of OCS. From this modeling, we extract $\frac{\partial e(\nu)}{\partial\nu} = 0.467$ V/MHz and $V_\text{noise} = 398$ nV, after demodulation and 60 dB of amplification. Using Eq. (\ref{detnoise}), the predicted one-second Allan deviation is $1.9\times 10^{-12}$, or a factor of $\sqrt{2}$ lower for the modified Allan deviation. We find excellent agreement with the results of \cite{Greenberg2024}, validating the correctness of our model and suggesting that other noise sources are below the detector SNR limit in this configuration.

\subsection{Instability from the intermodulation effect}

Local oscillator phase noise at even harmonics of the modulation frequency mixes into the error signal and introduces a further source of instability, known as intermodulation noise \cite{Audoin1991, Bahoura2003}. In the quasistatic limit ($\Omega \ll \Gamma$), this source of noise is known to be \cite{Audoin1991},

\begin{equation}
    \sigma_\text{qs} (1 \text{ s}) \approx \frac{\Omega}{\nu_0} \sqrt{S_\phi (2\Omega)}, \label{qssigma}
\end{equation}

\noindent where $S_\phi(f)$ is the phase noise power spectral density of the local oscillator in a 1 Hz bandwidth, at the Fourier frequency $f$. However, the configurations considered here strongly violate the quasistatic limit, $\Omega \gg \Gamma$, and discrepancies with Eq. \eqref{qssigma} have been observed in this regime \cite{Barillet1993}. To our knowledge, there is no treatment available in the literature for the nonquasistatic intermodulation noise associated with an absorptive signal. 

To calculate this, we added a phase noise term to \eqref{modfield},

\begin{equation}
    E(t) = E_0 e^{i\Phi(t)} e^{i(2\pi\nu t + \beta \sin 2\pi \Omega t)},
\end{equation}

\noindent where $\Phi(t) = \beta_\Phi \sin(2\pi f_n t + \xi)$, for a noise frequency $f_n$, noise amplitude $\beta_\Phi$, and phase $\xi$. Exploiting the Jacobi-Anger expansion and assuming $\beta_\Phi \ll 1$, we express this term as,

\begin{equation}
    e^{i \Phi(t)} \approx 1 + \frac{\beta_\Phi}{2} e^{i(2\pi f_n t + \xi)} - \frac{\beta_\Phi}{2} e^{-i(2\pi f_n t + \xi)}.
\end{equation}

\noindent Including phase noise transforms \eqref{eleceq} into

\begin{equation}
\begin{split}
    \frac{E_T(t)}{E_0 e^{2\pi i\nu t}} =& \sum_{m = -\infty}^\infty J_m(\beta) \left[ T(\nu + m \Omega) e^{2\pi i m \Omega t} \right. \\
    +& T(\nu + m\Omega+f_n)\frac{\beta_\Phi}{2}e^{2\pi i (m \Omega + f_n) t + i\xi} \\
    -& \left. T(\nu + m\Omega -f_n)\frac{\beta_\Phi}{2}e^{2\pi i (m \Omega - f_n) t - i\xi} \right].
\end{split}
\end{equation}

\noindent This equation was used to simulate the intermodulation effect, using the methods outlined above. The noise amplitude at $f_n$ is given by $\beta_\Phi = 2\sqrt{S_\phi (f_n)}$. Phase noise at even multiples of $\Omega$ contributes to the intermodulation noise, with $f_n = 2 \Omega$ by far the leading source. For modulation frequencies far outside the quasistatic regime, we find that, for the range of parameters of interest in this study, the instability is well approximated by

\begin{equation}
    \sigma_\text{nqs} \approx \frac{\Gamma}{2\nu_0} \sqrt{S_\phi (2\Omega)} \approx  \frac{\Gamma}{2\Omega} \sigma_\text{qs}.
\end{equation}

This source of instability is a key design consideration for the realization of practical THz frequency references. Using the measured phase noise of the dual wavelength Brillouin laser at 1 MHz \cite{Heffernan2024}, the intermodulation noise for the conditions of \cite{Greenberg2024} amounts to $\sigma_\text{IM} (1 \text{ s}) = 1.8\times10^{-13}$, an order of magnitude below the instability from detector SNR. For other frequency references, intermodulation noise can be dominating: for instance, one inexpensive, compact, and high-power THz local oscillator candidate is the resonant tunneling diode \cite{Asada2021}. However, this device has relatively deleterious phase noise. This can be partially ameliorated by high-frequency modulation: 10-MHz modulation can be realized by incorporating a varactor diode in the slot antenna of the RTD \cite{Ogino2017}. The phase noise at 20 MHz for an RTD has been measured to be -90 dBc/Hz \cite{Wang2015}, implying a one-second instability in the middle of the $10^{-11}$ decade, still well above the intermodulation noise of the Brillouin laser but potentially attractive for certain applications.

\subsection{Instability from other noise sources}

Photon shot noise contributes instability due to the Poissonian statistics of photon counting. However, for THz powers on the order of 10 $\upmu$W or higher, the photon shot noise-limited SNR is greater than $10^{8}$, several orders of magnitude greater than the SNR from current shot noise attributed to the THz detector, so the latter is the limiting factor.

Quantum projection noise arises due to the intrinsic randomness of quantum measurements \cite{Itano1993, Vershovskii2020}. For phase modulation spectroscopy, this instability is approximately,

\begin{equation}
    \sigma_y (1 \text{ s}) \approx \frac{1}{\nu_0}\sqrt{\frac{2 \Delta\nu_\text{tot}}{N F V}}.
\end{equation}

\noindent For the conditions in \cite{Greenberg2024}, there were $2.6 \times 10^{12}$ molecules of OCS in the desired rotational state ($J'' = 25$) in the 13-cm section of WR-3.4 waveguide in which the molecular resonance was probed. This leads to a contribution of $2\times 10^{-15}$ from quantum projection noise, orders of magnitude below the SNR limit. For all the experimental conditions investigated here, the quantum projection noise is well below the instability from other sources.

Intensity noise of the THz source at a Fourier frequency corresponding to the modulation rate can lead to instability in the locked signal, in a manner precisely analogous to the detector signal-to-noise ratio, described in Section II(B), above. Specifically, if such intensity noise induces root-mean-squared voltage fluctuations, $V_\text{noise}$, on the output of the detector, then Eq. (\ref{detnoise}) is valid for this source of instability as well. In the photomixed dual-wavelength Brillouin laser, residual intensity noise at Fourier frequencies on the MHz scale are about -140 dBc/Hz \cite{Heffernan2024b}, leading to an SNR of 70 dB on the error signal. For comparison, the detector-limited SNR is an order of magnitude greater \cite{Greenberg2024}. However, use of up-converted electronic or direct-THz local oscillators demands careful evaluation of this source of instability.

\subsection{Residual amplitude modulation}

The theoretical treatment so far has focused on the fundamental limitations to short-term instability arising from white FM noise sources, with instability decreasing as $\tau^{-1/2}$ for averaging time $\tau$. However, a fundamental technical limitation for phase modulation spectroscopy is non-white noise associated with residual amplitude modulation (RAM). This effect arises when a system intended to impart pure phase modulation instead introduces unwanted amplitude fluctuations, distorting the error signal and shifting the frequency lock point from its unperturbed location \cite{Bjorklund1983}. Since frequency stabilization in molecular references relies on precise detection of phase variations any variance of the RAM leads to systematic noise and instability. 
RAM has multiple origins depending on the method used to generate and modulate the terahertz signal. In systems employing electro-optic phase modulators RAM can result from residual birefringence due to stress-induced variations in the crystal's refractive index and misalignment of the input polarization relative to the modulator axis \cite{Wong1985}. Etalon effects from parasitic reflections within the modulator or other optical components further contribute to RAM by creating interference fringes that fluctuate with small environmental changes. When frequency modulation is applied directly to a terahertz source such as a resonant tunneling diode or quantum cascade laser RAM arises due to intrinsic amplitude to frequency coupling. In resonant tunneling diodes the bias voltage modulates both the carrier frequency and the output power causing small intensity fluctuations at the modulation frequency. Similarly in quantum cascade lasers RAM is linked to the laser’s current tuning response where frequency shifts are accompanied by variations in gain and output power.
The effect of RAM on frequency stability depends on its magnitude and how it interacts with the phase detection process. A primary consequence is the introduction of a parasitic component in the demodulated error signal shifting the apparent lock point of the frequency reference. This bias varies over time due to slow drifts ($>100$ ms) in RAM, resulting in frequency deviations. The impact on short term instability scales with the magnitude of the RAM and the modulation frequency itself, since aliasing effects fold amplitude noise into the phase lock loop. 
Mitigation strategies depend on the nature of the terahertz source and modulation scheme. In systems using electro-optic modulators, active cancellation techniques can apply a correction signal to suppress RAM dynamically \cite{Zhang2014}. Optimization of modulator bias and polarization alignment reduces birefringence-induced RAM. In directly modulated sources such as resonant tunneling diodes and quantum cascade lasers bias optimization and external amplitude stabilization methods can minimize the effect. In any implementation avoiding etalon effects through the use of wedged optics and anti-reflection coatings reduces RAM at the source \cite{Bi2019}.
RAM remains a fundamental limitation in terahertz molecular frequency references across all modulation schemes. Understanding its origin and actively suppressing it is necessary for achieving optimal instability on longer timescales.


\section{Optimal instability for various gases}

\begin{figure}[!t]
\centerline{\includegraphics[width=\columnwidth]{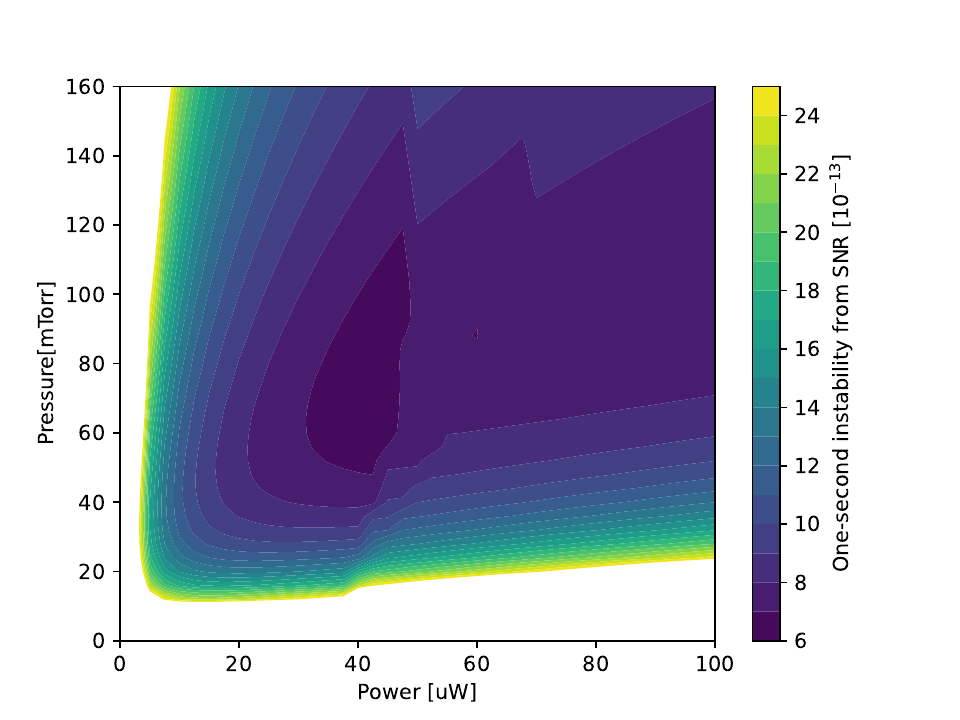}}
\caption{By adjusting experimental parameters in the simulation, the SNR instability for a certain gas could be optimized. In this plot, the optimal conditions for OCS in waveguide was found to be at 40 $\upmu$W and 70 mTorr.}
\label{fig3}
\end{figure}

A number of commercially available gas molecules have attractive qualities for operation as a THz frequency standard, and selecting the most optimal candidate is by no means trivial. Necessary considerations include the molecule's mass, dipole moment, molecular structure (linear, symmetric top, or asymmetric top), rotational constant, hyperfine structure, as well as extrinsic experimental properties such as pressure, power, and path length. By the theory here adumbrated, we can determine optimized operational conditions for each molecule and compare the attainable instabilities under a variety of conditions. Due to substantial covariance among the parameters, it was necessary to iterate the optimization protocols until convergence was achieved. As an example, Fig. \ref{fig3} illustrates the calculated instability as a function of pressure and power for OCS in waveguide.

\begin{table}
\caption{Optimized parameters for spectroscopy in waveguide}
\label{table}
\setlength{\tabcolsep}{3pt}
\begin{tabular}{p{36pt}p{40pt}p{46pt}p{35pt}p{35pt}p{30pt}}
\hline
Molecule& Frequency [GHz]& Pressure [mTorr]& Power [$\upmu$W]& Length [cm]& $\sigma_\text{SNR}$(1s) [$10^{-13}$]\\
\hline\hline
OCS& 316& 70& 40& 35& 6.5\\
\hline
N$_2$O& 301& 150& 1300& 75& 18\\
\hline
HC${^{15}\text{N}}$& 258& 35& 25& 20& 17\\
\hline
CH$_3$CN& 313& 50& 65& 50& 36\\
\hline
OCSe& 313& 80& 45& 40& 7.5\\
\hline
CO& 231& 140& 2200& 120& 42\\
\hline
CH$_3$F& 306& 80& 50& 35& 16\\
\hline
HI& 385& 100& 75& 40& 5.3\\

\hline\hline
\multicolumn{6}{p{251pt}}{Power is quoted as emitted power, before waveguide losses and absorption. The isotope employed is that with the greatest natural abundance, unless stated otherwise.}
\end{tabular}
\label{tab1}
\end{table}

First, we examine the instability for a variety of gases in waveguide, with frequencies near the water window at 300 GHz. The results of this analysis are shown in Table \ref{tab1}. With the exception of HI, it is appropriate to use WR-3.4 waveguide with these molecules, which exhibits losses typically close to 0.2 dB/cm. From the simulations, it is clear that higher modulation frequency leads to decreased instability as a result of reduced $1/f$ noise on the detector. For this reason, a relatively high modulation frequency of 100 MHz has been assumed throughout, though depending on the THz LO being employed, attaining such a high modulation frequency may not be straightforward.

For the gases analyzed, it is assumed that the isotopologue employed will be that with the greatest natural abundance unless stated otherwise. Isotopes with nuclear spin $I\geq 1$ contribute quadrupolar hyperfine structure. Among the atoms on Table \ref{tab1}, $^{14}$N and $^{127}$I have quadrupolar nuclear structure. For $^{14}$N$_2$O and CH$_3$C$^{14}$N, quadrupolar hyperfine structure leads to kHz-scale, unresolved splitting between the strong $\Delta F = +1$ lines, with minimal consequences to the operation of the frequency reference. By contrast, HC$^{14}$N has three strong lines within 300 kHz of each other, leading to very strong, pressure- and temperature-dependent line pulling. For this reason, we suggest that the isotopologue HC$^{15}$N is a more appropriate frequency reference, since $^{15}$N lacks quadrupolar structure. The three hyperfine lines of the H$^{127}$I, $J' = 1\leftarrow J'' = 0$ transition are well resolved, by 200 MHz, so the effect of the hyperfine splitting is just to moderately decrease the line intensity. Additionally, isotopes with nuclear spin $I = 1/2$ have magnetic hyperfine structure, though this is much smaller than quadrupolar nuclear structure and in practice has minimal implications for operation in any of the gases analyzed.

Stabilizing to gases in free-space could have several marked advantages, including lower propagation losses, decreased cost, and decreased broadening from intensity and wall collisions. By increasing the path length, the absorption signal can be increased, allowing for decreased pressures and narrower lines; with a long enough path, the linewidth of the transition can be limited almost entirely by Doppler broadening, implying that the heaviest atoms will have the lowest instabilities. Investigation of Doppler-free probing schemes is beyond the scope of this paper. In Table \ref{tab2}, we list the optimal instability that can be achieved for a variety of path lengths. For reasonable experimental parameters, free space spectroscopic setups stand to realize instability an order of magnitude below waveguide schemes. On the other hand, etalons have the possibility of presenting a serious limitation for free space frequency stabilization, though the recent development of low-loss free space THz isolators could mitigate this \cite{Grebenchukov2022}. 

\begin{table}
\caption{Instability for free space setups of variable path length}
\label{table}
\setlength{\tabcolsep}{3pt}
\begin{tabular}{p{38pt}p{35pt}p{35pt}p{35pt}p{35pt}p{35pt}}
\hline
Molecule& $\sigma_\text{SNR}$ at: \\
& 10 cm& 50 cm& 1 m& 10 m& $\infty$\\
\hline\hline
OCS& 2.9& 0.82& 0.69& 0.66& 0.65\\
\hline
N$_2$O& 43& 8.6& 4.3& 0.82& 0.75\\
\hline
HC${^{15}\text{N}}$& 1.9& 1.0& 0.99& 0.97& 0.97 \\
\hline
CH$_3$CN& 5.3& 1.4& 1.0& 0.85& 0.84\\
\hline
OCSe& 3.0& 0.73& 0.55& 0.48& 0.48\\
\hline
CO& 220& 44& 22& 2.3& 0.95\\
\hline
CH$_3$F& 4.9& 1.2& 0.97& 0.87& 0.86\\
\hline
HI& 3.0& 0.71& 0.51& 0.44& 0.44\\

\hline\hline
\multicolumn{6}{p{251pt}}{All instabilities are fractional and in units of $10^{-13}$, evaluated at an averaging time of 1 s. The column labeled $\infty$ shows the limit as the path length becomes large. For compactness, the THz power and gas pressure is not listed on this table, though the optimal power is always on the order of 250 $\upmu$W, and the optimal pressure varies, depending on the gas and path length, from below 1 mTorr to several 100 mTorr.}
\end{tabular}
\label{tab2}
\end{table}

\section{Conclusions}
\label{sec:guidelines}

We have carried out an analysis of THz frequency references stabilized to rotational transitions in a variety of commercially available molecules, finding optimized operational conditions for each. We find that the instability of such frequency references can be limited by detector SNR. We note that continuing development and evaluation of low-noise THz detectors, for instance Fermi-level managed barrier diodes \cite{Ito2015} or superconducting hot-electron bolometers \cite{Shurakov2016}, is a promising direction forward for the improvement of THz frequency references. Making use of the detector explored in this work, molecules confined to waveguide have the prospects of achieving a one-second instability below the $10^{-12}$ decade. In free space, instability an order of magnitude lower is attainable. Gases with very large dipole moments, such as HCN, can exhibit superior instability for very short path lengths (see Table \ref{table}), and they could be preferred for compact systems. On the other hand, if space is not a critical consideration, it is found that HI, OCSe, and OCS exhibit exceptional performance. However, it must be noted that the high transition frequency (385 GHz) of the fundamental transition of HI is likely to present technical challenges in implementation, and that OCSe is not industrially manufactured. OCS emerges from this analysis as the most readily exploitable molecule for such frequency references. 

\appendix

\begin{figure}[!t]
\centerline{\includegraphics[width=\columnwidth]{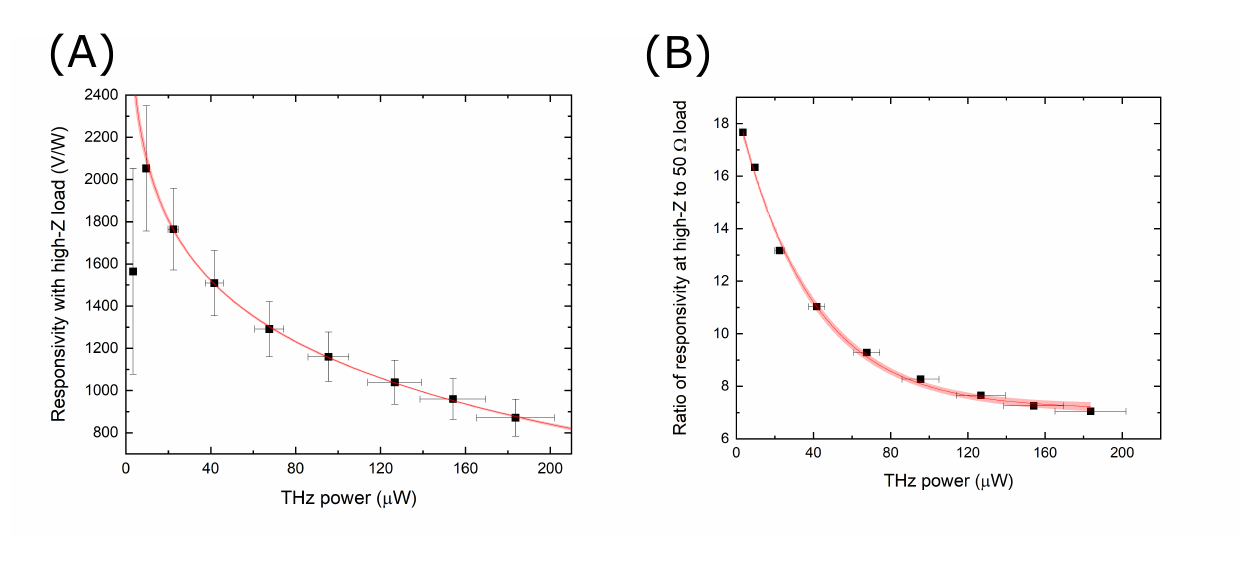}}
\caption{To construct a detector model, the output of the detector was measured as a function of THz power. (A) With a high-Z load, the detector is found to depart from square-law operation at a few 10 $\upmu$W, with responsivity gradually decreasing as the power increases to several 100 $\upmu$W. The data is fit to a function of form $\text{a}+\text{b} x^\text{c}$. (B) with a 50-$\Upomega$ load, the voltage output is reduced due to the current-limiting nature of the diode, and this reduction factor changes as a function of THz power. An exponential fit is found to model the variation of this ratio well. All error bars are the 1-$\upsigma$ uncertainties, and the shaded region is the 68\% confidence interval of the fit function.}
\label{appfig1}
\end{figure}

To accurately simulate the response of the zero-biased SBD to THz radiation impinging upon it, it is necessary to construct a detector model. Fig. \ref{appfig1}(A) shows the detector response, measured in V/W, as a function of changing THz power. The THz radiation is generated by using a UTC-PD to photomix two optical lines. The UTC-PD photocurrent measures THz power, by means of a manufacturer calibration, though this measure becomes inaccurate at low powers. The data are fit well by a function of the form $\text{a}+\text{b} x^\text{c}$. Below a few 10 $\upmu$W, the responsivity is expected to be constant, though our measurement is not very sensitive in this range. For this reason, we model the responsivity as constant at 2000 V/W for low THz powers. These measurements were taken with a high impedance (high-Z) load, but during locked operation in \cite{Greenberg2024}, the output of the SBD was passed through a bias-tee with a 50-$\Upomega$ load on the DC output. Due to the current-limiting nature of the diode, the output is reduced by a factor which is variable with THz power, as shown in Fig. $\ref{appfig1}$(B). This reduction factor was found to vary exponentially with power.

The $1/f$ noise of the SBD also varies with THz power, increasing approximately linearly over the range studied. At a Fourier frequency of 1 Hz, the noise increases from 5.3 $\upmu$V/$\sqrt\text{Hz}$ at a power of 22 $\upmu$W to 9.8 $\upmu$V/$\sqrt\text{Hz}$ at a power of 180 $\upmu$W. These values were measured with a high-Z impedance and are decreased by the same factor as the responsivity if a 50-$\Upomega$ load is applied. The $1/f$ noise at the modulation frequency is then added in quadrature with the thermal and shot noise, given by Eq. (\ref{thermshoteq}), to give the total voltage fluctuation, $V_\text{noise}$, from detector SNR. This quantity is used in Eq. (\ref{detnoise}) to estimate the total instability arising from detector SNR.





\bibliography{mybib}{}
\bibliographystyle{ieeetr}
\end{document}